\documentclass[onecolumn,notitlepage,superscriptaddress]{revtex4-1}

\usepackage{amsmath}
\usepackage{paralist}
\usepackage{listings}
\usepackage{url}
\usepackage{hyperref}
\usepackage[utf8]{inputenc}

\begin{document}

\title{RMPCDMD: Simulations of colloids with coarse-grained hydrodynamics, chemical
  reactions and external fields}
\author{Pierre de Buyl}
\affiliation{Instituut voor Theoretische Fysica, KU Leuven B-3001, Belgium}
\author{Mu-Jie Huang}
\affiliation{Chemical Physics Theory Group, Department of Chemistry, University of Toronto, Toronto, Ontario M5S 3H6, Canada}
\author{Laurens Deprez}
\affiliation{Instituut voor Theoretische Fysica, KU Leuven B-3001, Belgium}

\begin{abstract}
The RMPCDMD software package performs hybrid Molecular Dynamics simulations, coupling
Multiparticle Collision Dynamics to model the solvent and Molecular Dynamics to model
suspended colloids, including hydrodynamics, thermal fluctuations, and chemically active
solvent particles and catalytic colloids.
The main usage of RMPCDMD is the simulation of chemically powered nanomotors, but other
setups are considered: colloids in the presence of a thermal gradients or forced flows.
RMPCDMD is developed in Fortran 2008 with OpenMP for multithreaded operation
and uses the HDF5-based H5MD file format for storing data.
RMPCDMD comes with documentation and a tutorial for the simulation of chemically powered
nanomotors.
\end{abstract}
\maketitle

\section*{Keywords}

Molecular Dynamics ; Colloids ; Hydrodynamics ; Nanomotors ; Fortran ; Python

\section{Overview}

\subsection{Introduction}

Molecular Dynamics (MD) is a computational technique to study condensed matter systems by
numerically solving Newton's equations of motion for a large number of
particles~\cite{frenkel_smit_2002}.
Often, due to the number of constituents in a system, so-called {\em coarse-grained} or
{\em multiscale} strategies are used to reduce the computational cost of a simulation.

One simulation strategy for modelling colloids embedded in a solvent is to
couple the Multiparticle Collision Dynamics (MPCD) algorithm~\cite{malevanets_kapral_mpcd_1999}
for the solvent with MD for the colloids~\cite{malevanets_kapral_mpcd_2000}. It differs from {\em
  traditional} MD by the use of a coarse-grained representation of the solvent's structure
and dynamics and by the possible occurrence of chemical species transformations in the course
of the simulation.
In the latter case, the extension Reactive MPCD (RMPCD) provides bulk chemical kinetics that
reduce to reaction--diffusion equations~\cite{rohlf_et_al_rmpcd_2008}.
The name RMPCDMD results from the concatenation of the different algorithms' acronyms:
Reactive Multiparticle Collision Dynamics and Molecular Dynamics.

The focus of RMPCDMD lies on the modeling of chemically active colloids
(e.g. nanomotors~\cite{ruckner_kapral_prl_2007}) or of colloids subjected to external
gradients (e.g. flows~\cite{prohm_inertial_2012} or temperature
gradients~\cite{lusebrink_thermophoresis_2012}).
A RMPCDMD simulation consists in $N_\mathrm{solvent}$ solvent particles and
$N_\mathrm{colloids}$ colloidal particles with a mass, position, and velocity,
placed in a cuboid domain.
Colloid--solvent and colloid--colloid interactions follow Lennard--Jones type
potentials, but there is no interaction potential acting among the solvent
particles. As $N_\mathrm{solvent} \gg N_\mathrm{colloids}$, this considerably
speeds up a simulation in comparison to a full MD simulation.
The time evolution of colloidal and solvent particles follows Newton's equations
of motion, which are solved iteratively using the velocity-Verlet
algorithm~\cite{allen_tildesley_1987,frenkel_smit_2002}.
The output from a simulation is a trajectory representing the coordinates of all particles
in the course of time.
Typically, solvent data is preaveraged during the simulation as its storage requires
significant amounts of space.

\subsection{Implementation and architecture}

\subsubsection{Development}

RMPCDMD is implemented in Fortran, following the 2008 version of the standard, and uses
OpenMP for multi-core execution.
Trajectories are stored in the H5MD file format~\cite{h5md_cpc_2014}, and Python programs are
provided to illustrate data analysis.
The module \texttt{fortran\_h5md}~\cite{fortran_h5md_github} provides helper routines to write H5MD
files from Fortran.
H5MD files are HDF5 files (\url{https://www.hdfgroup.org/HDF5/}) with an internal
organization tailored for molecular simulations~\cite{h5md_cpc_2014}.

In designing RMPCDMD, we tried to adopt practices for scientific computing that help to
deliver reproducible results and facilitate the extension of the code and its maintenance.
We also made special efforts on the usability of the software.
The design choices are the results of the authors' experience on scientific programming,
with further motivation by the article on ``Best Practices for Scientific Computing'' by
Wilson~{\em et al}~\cite{best_practices_2014}:
\begin{itemize}
\item The algorithms are implemented in Fortran modules in the \texttt{src/} directory and
shared between the simulations found in \texttt{programs/}. Table~\ref{tab:src} summarizes
their content.
\item Subroutines and functions are limited in size, so that understanding any single
code unit remains reasonable.
\item Data is encapsulated in {\em derived types}.
\item No global variable is used, to clarify to the reader what data is passed to subroutines.
\item The programming style strives for legibility, conciseness, and efficiency.
\item All simulation parameters are given from a text configuration file, there is no
hard-coded setting in the code that would require re-compilation between runs.
\item The code is tracked with the git revision control software~\cite{git_web}
and published on GitHub (\url{https://github.com/}).
\item The build procedure and partial testing are automatically executed on the Travis CI
platform (\url{https://travis-ci.org/}) for continuous integration.
\end{itemize}

\begin{table}[h]
\centering
\begin{tabular}{l p{14cm}}
  \hline
  Module name & Functionality\\
  \hline
  \hline
  \texttt{common} & Routines of general use: histogramming, parsing of command-line arguments, timing and minimum-image convention function.\\
  \texttt{hilbert} & Computation of the compact Hilbert index of a lattice cell and vice-versa.\\
  \texttt{cell\_system} & Cell data structure.\\
  \texttt{particle\_system} & Particles data structure and sorting routine.\\
  \texttt{particle\_system\_io} & Data structures to facilitate the storage of thermodynamic and particle data.\\
  \texttt{interaction} & Computation of Lennard-Jones force and energy.\\
  \texttt{neighbor\_list} & Data structure and routines to update the neighbor list.\\
  \texttt{mpcd} & Algorithms for MPCD streaming and collision, and for bulk reactions.\\
  \texttt{md} & Algorithms for velocity-Verlet and rigid body MD.\\
  \hline
\end{tabular}
\caption{A summary of the functionality provided by the Fortran modules of RMPCDMD found in the
\texttt{src/} directory.}
\label{tab:src}
\end{table}
Instead of building a generic program that could handle the logic of different geometries,
boundary conditions, and colloid types, specific simulation programs have been
implemented.
Each program calls the algorithms from the modules and resembles a statically compiled
``simulation script''. The amount of administrative code (reading parameters, defining
storage, etc.) is however larger than what could be achieved in a higher-level language
such as Python.

CMake is used to build RMPCDMD and run automated tests in \texttt{test/} that cover
some of the base algorithms such as particle sorting or the histogram routine.
The Fortran module \texttt{fortran\_tester}~\cite{fortran_tester_github} provide convenience
routines for testing (e.g. equality assertions).
Automated testing using actual simulations is difficult and very much time
intensive, and we rely on it only for a small part of the code.
In simulations without a thermostat or wall interactions, the total momentum and energy of
the system must be conserved, which is verified for a short run of the
\texttt{single\_dimer\_pbc} simulation as part of the continuous integration testing.
Further validation of the code is done via comparison of physical results: the velocity
distribution of particles in equilibrium, the viscosity of the fluid, etc.

\subsubsection{User perspective}

For the end-user, the interface comes as a single command-line tool, \texttt{rmpcdmd}, that
drives the simulations and provides access to utility programs. A typical command is
\begin{lstlisting}[language=bash]
rmpcdmd run single_dimer_pbc dimer.parameters dimer.h5 auto
\end{lstlisting}
where \texttt{single\_dimer\_pbc} is the simulation program, \texttt{dimer.parameters} is
the configuration file, \texttt{dimer.h5} the output datafile, and \texttt{auto} is the
setting for the random-number seed.
While the individual programs can be run directly, the command-line tool \texttt{rmpcdmd}
aims to provide a single file to place in the program ``PATH'' on the
user's computer and to automate some actions.
The commands available are:
\begin{compactdesc}
\item[\texttt{run}] execute one of the simulation programs. The setting of the environment
variable \texttt{OMP\_NUM\_THREADS}, and the start and end time of the run are shown. Bash's
\texttt{time} provides the ``user time'' actually used by the program.
\item[\texttt{seeder}] print out a signed 64-bit integer value generated by the computer's
\texttt{/dev/urandom} device.
\item[\texttt{plot}] display an observable from a simulation file.
\item[\texttt{timers}] print out (or display) the timing information from a simulation file.
\end{compactdesc}

A benefit of using separate simulation programs is that the configuration file for a
simulation only contains directly relevant parameters.
All simulation programs share the same command-line syntax, read parameters from a text file
using the ParseText Fortran library~\cite{ParseText_github}, read the seed for the RNG from
the command line and output the trajectory to a H5MD file.
An example configuration file for \texttt{single\_dimer\_pbc} is listed in Appendix~\ref{sec:example}.

A generic plotting utility is provided and can be invoked as
\begin{verbatim}
rmpcdmd plot dimer.h5 --obs temperature
\end{verbatim}
to plot the temperature, for instance.
This plotting program and the other analysis programs that come with RMPCDMD are written in
Python and rely on NumPy for storing and processing numerical data, SciPy~\cite{scipy-web}
for numerical algorithms, Matplotlib~\cite{matplotlib_2007} to produce 2D figures,
h5py~\cite{collette_python_hdf5_2014} to read HDF5 files and Mayavi~\cite{mayavi_2011} to
visualize 3D data.

\subsection{Documentation}

A documentation is found in the directory \texttt{doc/}. It is published on the web,
\url{http://lab.pdebuyl.be/rmpcdmd/} and also serves as the homepage of RMPCDMD.
The documentation is built with the Sphinx documentation generator~\cite{sphinx_doc_web} and
is written in reStructuredText.
An extension to Sphinx is bundled with RMPCDMD to provide (i) automatic links from the
documentation to the source code documentation that is annotated with
Doxygen (\url{http://www.stack.nl/~dimitri/doxygen/}) and (ii) automatic inclusion
of the Doxygen header of Fortran programs.
Specific instructions for the installation and execution of RMPCDMD are provided, as well as
general informations on the algorithms are presented.
The description of the simulation programs is generated automatically from the beginning of
the corresponding \texttt{.f90} files and includes all parameters for the simulations.

A tutorial on the simulation of chemically powered nanomotors, based on RMPCMD, was given at
the recent workshop ``Modeling of chemically powered nanomotors'' held at the KU Leuven in
April 2016~(\url{http://lab.pdebuyl.be/2016-nanomotor/}). This tutorial is
reproduced in the documentation of RMPCDMD and represents a unique resource in the field of
nanomotor simulation.

\subsection{Algorithms}

For completeness, we state here exactly what algorithms are implemented in RMPCDMD, with the
corresponding reference to the literature.

RMPCDMD implements the original MPCD collision rule~\cite{malevanets_kapral_mpcd_1999} and
the Anderson thermostat~\cite{noguchi_epl_2007}.
Random shifting of the spatial grid is always performed~\cite{ihle_kroll_srd_2001} to ensure
Galilean invariance.
Flows are generated by applying a constant acceleration to solvent
particles~\cite{allahyarov_gompper_mpcd_flows_2002,whitmer_luitjen_2010}.
Stick boundary conditions at the wall use the bounce-back rule and virtual particles during
the collision step~\cite{lamura_mpcd_epl_2001}.

Molecular Dynamics is performed with the velocity-Verlet
algorithm~\cite{allen_tildesley_1987,malevanets_kapral_mpcd_2000,frenkel_smit_2002}.
The interaction potential, both for solvent--colloid and colloid--colloid interactions, are
purely repulsive cut-off 12--6 Lennard-Jones potentials of the form
\begin{equation*}
V(r) = 4\epsilon \left( \left(\frac{\sigma}{r}\right)^{12} - \left(\frac{\sigma}{r}\right)^{6} + \frac{1}{4} \right) \textrm{ for } r < 2^{1/6} \sigma
\end{equation*}
where the energy parameter $\epsilon$ and the scale parameter $\sigma$ depend on the species
considered.
Verlet and cell lists are used to compute the solvent--colloid forces and
energies~\cite{frenkel_smit_2002}. The sorting of solvent particles in cells is the same as
for the MPCD collision step.
The rigid body dynamics uses RATTLE~\cite{andersen_rattle_1983} in its iterative
implementation for the Janus particle and exactly for the dimer motor.

Spatial sorting of the solvent particles is based on compact Hilbert
indices~\cite{hamilton_compact_hilbert_tr}.
Random numbers are generated with the Threefry 2x64 RNG~\cite{random123}, allowing each
OpenMP thread to generate random numbers independently.
The RNG is provided by the \texttt{random\_module} library~\cite{random_module_github},
where it is implemented in C99.
The choices are inspired by the
\texttt{nano-dimer}~\cite{colberg_nanodimer_web} simulation code, which
implements these algorithms in OpenCL C for large-scale parallel processors.

To obtain a consistent measure of the temperature $T$ even in the presence of flows, we
compute $T$ by averaging the center-of-mass reference frame temperature over all MPCD cells
as
\begin{equation*}
T = \frac{1}{3N_c} \sum_\xi \frac{1}{N_\xi-1} \sum_i m_i \left( v_i - v_\xi \right)^2
\end{equation*}
where $N_c$ is the number of cells, the variable $\xi$ represents a cell, $N_\xi$ the number
of particles in a cell, $v_\xi$ the center-of-mass velocity of the cell, and $m_i$ and $v_i$
the mass and velocity of particle.

\subsection{Applications of RMPCDMD}

The prototypical use of the MPCD algorithm to study chemically active particles is the dimer
nanomotor, whose model was introduced by Rückner and Kapral~\cite{ruckner_kapral_prl_2007}.
It consists of two rigidly linked spheres evolving in a solvent.
Chemical reactions are catalyzed by one of the spheres and create a local concentration
gradient, effectively propelling the dimer along its symmetry axis.
This simulation is implemented in the program \texttt{single\_dimer\_pbc}.

To facilitate usage of RMPCDMD by newcomers, ready-made simulation ``experiments'' are
provided in \texttt{experiments/}. There, makefiles allow the user to simply type
\begin{verbatim}
make simulation
\end{verbatim}
to start a simulation.
The terminology that we have chosen reflects our opinion that computer simulations can be
viewed as numerical experiments in which one prepares a setup and conducts a study to
understand a certain physical phenomenon.
The ability to modify a given experimental choice (i.e. thermostating or not, surface
properties, etc.) to assess its effect is therefore critical. By providing several chemical
models and collision rules, RMPCDMD provides such a framework.

Four programs are provided to analyze the data of this experiment.
\texttt{plot\_velocity.py} displays the laboratory reference frame velocity or the directed
velocity of the dimer (with the option \texttt{--directed}).
\texttt{plot\_histogram.py} displays the cylindrical shell histogram of the reaction
product concentration.
\texttt{plot\_msd.py} displays the mean-squared displacement of the dimer's center-of-mass
position. This quantity is a practical interest in comparison to experimental
studies~\cite{howse_et_al_prl_2007}.
\texttt{view\_last\_simulation\_frame.py} displays a 3D representation of the dimer motor
using the Mayavi visualization library~\cite{mayavi_2011}, either with its full trajectory
as a line (with option \texttt{--unwrap}) or with the product solvent particles B (with option
\texttt{--show-B}).
The two first analysis programs are modeled after the results by Rückner and
Kapral~\cite{ruckner_kapral_prl_2007}.

While the aims of \texttt{single\_dimer\_pbc} are to reproduce
Ref.~\cite{ruckner_kapral_prl_2007} and to provide a starting point for other developments,
the other simulation programs in RMPCDMD are related to forward-looking research projects
that will benefit from a robust codebase. These other setups are:
\texttt{poiseuille\_flow} generates a Poiseuille flow between parallel plates, with no colloid.
\texttt{chemotactic\_cell} generates a Poiseuille flow with a two-species chemical gradient
perpendicular to the flow and an embedded spherical colloid or a dimer nanomotor.
\texttt{single\_janus\_pbc} implements a composite Janus nanomotor~\cite{de_buyl_kapral_nanoscale_2013}.
\texttt{single\_sphere\_thermo\_trap} generates a thermal gradient between two plates, with
an embedded spherical colloid.

\section{Availability}

\begin{description}
\item[Operating System] RMPCDMD has been tested on Linux using the gcc/gfortran and
icc/ifort compilers and on OS X (El Capitan) using gcc/gfortran (installed via MacPorts).
\item[Programming Language] Fortran 2008 for the simulation code, C99 for the Random Number
Generator, bash and make for the execution and Python (with NumPy, SciPy, matplotlib and
h5py) for analysis.
\item[Additional system requirements] The largest RAM usage in RMPCDMD comes from the
solvent coordinates and requires about $N_\textrm{solvent}\times$250~bytes of storage. The
example simulations in the \texttt{experiments/} directory require between 100MB and 200MB
or RAM.
\item[Dependencies] The CMake and Make tools, and the HDF5 library. For data analysis,
NumPy, SciPy, matplotlib and h5py.
\end{description}

\section*{List of contributors}

\begin{compactdesc}
\item[Pierre de Buyl] Project creator, code, documentation, build system, tutorial.
\item[Laurens Deprez] Code: RATTLE for the dimer, angle-dependent MPCD, boundary conditions
and forcing for flows, Lennard-Jones 9-3 walls, chemotactic cell setup.
\item[Mu-Jie Huang] Tutorial: Janus particle.
\item[Peter Colberg] Initial OpenMP parallelization, continuous integration testing and
build system.
\end{compactdesc}

\section*{Software location}

\subsection*{Archive}

\begin{compactdesc}
\item[Name] Zenodo
\item[Persistent identifier] DOI
\item[Licence] BSD 3-clause
\item[Publisher] Pierre de Buyl
\item[Version published] 0.1
\item[Date published] TODO
\end{compactdesc}

\subsection*{Code repository}

\begin{compactdesc}
\item[Name] GitHub
\item[Persistent identifier] \url{https://github.com/pdebuyl-lab/RMPCDMD}
\item[Licence] BSD 3-clause
\item[Date published] TODO
\end{compactdesc}

\section{Reuse potential}

Besides \texttt{nano-dimer}~\cite{colberg_nanodimer_web} (by Peter Colberg, available under
the MIT license), RMPCDMD is the only open-source software to provide support for
chemically-powered nanomotor simulation.
RMPCDMD has, in addition to \texttt{nano-dimer}, support for external fields or rigid-body dynamics
for assemblies larger than the dimer.
\texttt{nano-dimer}, on the other hand, is based on OpenCL C and on the Message Passing
Interface for distributed parallel execution, and is able to handle many-motors simulations
that are out of reach of RMPCDMD for performance reasons.
Given the research activity on the topic of active colloids, we believe that providing a
reference implementation of a chemically active MPCD fluid can serve as a starting point for
further modeling, by us or by other groups.

RMPCMD has not been used for published work yet. It is in use for ongoing research projects
at the {\em Instituut voor Theoretische Fysica} of the KU Leuven.

\section*{Acknowledgments}

The authors thank Raymond Kapral for discussions and for motivating this line of research,
and Peter Colberg for his contributions to the code, his comments on the manuscript and for
interesting discussions.
The authors thank all authors of the open-source projects that they rely on for their
research.
PdB is a postdoctoral fellow of the Research Foundation-Flanders (FWO).

\section*{Competing interests}

The authors declare that they have no competing interests.

\bibliographystyle{jors}
\bibliography{rmpcdmd}

\appendix

\section{Example configuration file}
\label{sec:example}

An example configuration file for the program \texttt{single\_dimer\_pbc} is shown below.

\begin{lstlisting}
# physical parameters
T = 1
L = 32 32 32
rho = 10 
tau = 1
probability = 1.0
bulk_rmpcd = T
bulk_rate = 0.1

# simulation parameters
N_MD = 100
N_loop = 500

# interaction parameters
sigma_N = 3.0
sigma_C = 3.0

d = 6.5
epsilon_N = 1.0 0.2
epsilon_C = 1.0 1.0

epsilon_N_N = 1.0
epsilon_N_C = 1.0
epsilon_C_C = 1.0
\end{lstlisting}

\end{document}